\documentclass[%
 aip,
 amsmath,amssymb,
 reprint,%
]{revtex4-1}

\usepackage{graphicx}
\usepackage{dcolumn}
\usepackage{bm}

\usepackage[utf8]{inputenc}
\usepackage[T1]{fontenc}
\usepackage{mathptmx}
\usepackage{etoolbox}
\usepackage[table,xcdraw]{xcolor}
\usepackage{balance}
\usepackage{booktabs}
\usepackage{subfig}
\usepackage{comment}
\usepackage{multirow}
\newcommand{\Fig}[1]{Fig.\,{\ref{#1}}}
\newcommand{\Figs}[1]{Figs.\,{\ref{#1}}}

\makeatletter
\def\@email#1#2{%
 \endgroup
 \patchcmd{\titleblock@produce}
  {\frontmatter@RRAPformat}
  {\frontmatter@RRAPformat{\produce@RRAP{*#1\href{mailto:#2}{#2}}}\frontmatter@RRAPformat}
  {}{}
}%
\makeatother
\begin{document}

\preprint{AIP/123-QED}

\title[]{Time-periodic Metallic Metamaterials defined by Floquet Circuits}
\author{S. Moreno-Rodríguez}
\affiliation{ 
Department of Signal Theory, Telematics and Communications, Universidad de Granada (CITIC-UGR), 18071 Granada, Spain
}%

\author{A. Alex-Amor}%
\affiliation{%
Department of Information Technologies, Universidad San Pablo-CEU, CEU Universities,   Campus Montepríncipe,  28668 Boadilla del Monte (Madrid), Spain
}%

\author{P. Padilla}
\affiliation{ 
Department of Signal Theory, Telematics and Communications, Universidad de Granada (CITIC-UGR), 18071 Granada, Spain
}%

\author{J.F. Valenzuela-Valdés}
\affiliation{ 
Department of Signal Theory, Telematics and Communications, Universidad de Granada (CITIC-UGR), 18071 Granada, Spain
}%

\author{C. Molero}
\email{cmoleroj@ugr.es}
\affiliation{ 
Department of Signal Theory, Telematics and Communications, Universidad de Granada (CITIC-UGR), 18071 Granada, Spain
}%


\begin{abstract}
In this Letter, we study the scattering and diffraction phenomena in time-modulated metamaterials of metallic nature by means of Floquet equivalent circuits. Concretely, we focus on a time-periodic screen that alternates between ``metal" and ``air" states. We generalize our previous approaches by introducing the concepts of ``macroperiod" and ``duty cycle" to the time modulation. This allows to analyze time-periodic metallic metamaterials whose modulation ratios are, in general, rational numbers. Furthermore, with the introduction of the duty cycle, perfect temporal symmetry  is broken within the time modulation  as the time screen could remain a different amount of time in metal and air states. Previous statements lead to an enrichment of the diffraction phenomenon and to new degrees of freedom that can be exploited in engineering to control the reflection and transmission of electromagnetic waves. Finally, we present some analytical results that are validated with a self-implemented finite-difference time-domain (FDTD) approach. Results show that the scattering level and diffraction angles can be controlled independently by means of the duty cycle and the modulation ratio, respectively. Thus, novel time-based pulsed sources and beamformers can be efficiently designed.

\end{abstract}

\maketitle


The resolution of electromagnetic problems based on periodic structures has classically benefited from systematic simplifications thanks to the use of Floquet's theorem \cite{Elachi76, Alex3D_2022}. That is, the reduction of the complexity of the whole structure to a waveguide problem \cite{Varela2012}. Circuit models have proven to be very efficient tools to emulate waveguide environments \cite{Marcuvitz86, Costa2012, Mesa2018}. Simple models avoid the \emph{dynamic} behavior of the structure, combining transmission lines and quasi-static elements \cite{Luukkonen2008}. More sophisticated proposals include the contribution of higher-order modes/harmonics \cite{Berral2015, Alex2021, FloquetCircuit_2D2}. This implies the validity of the models for scenarios where higher-order harmonics have a leading role \cite{Molero2019}. This scenario is, for instance, quite common in time-varying systems, or in a more general context, in spacetime structures \cite{Taravati2019}.
 
Spacetime systems introduce time, generally in the form of a periodic modulation, as a new degree of freedom \cite{Pacheco2022, Caloz2020_1, Caloz2020_2, Deshmukh2022, Galiffi2022}. Though pioneering studies were theoretically reported in the middle of last century \cite{Morgenthaler1958, Tamir1964}, they have regained interest in the recent years, especially when non-reciprocity \cite{Zang2019, Sounas2021} was sought as a substitute of magnetic materials for insulators \cite{taravati2017}. Some other impressive properties have since then been reported, as new temporal mechanisms for amplification \cite{Pendry2021}, subharmonic mixing \cite{Wu2020}, giant bianosotropy \cite{Huidobro2021}, negative refraction \cite{Bruno2020} or an equivalent of the Brewster angle \cite{Pacheco2021Br}. Novel applications, just to name a few,  are proposed in the propagation domain focused on DOA estimation \cite{Fang2022}, imaging \cite{Kolner2020}, or digital processing \cite{Li2022}. 

\begin{figure}[!t]                  
        \centering	                
        \subfloat[]{\includegraphics[width=0.25\textwidth]{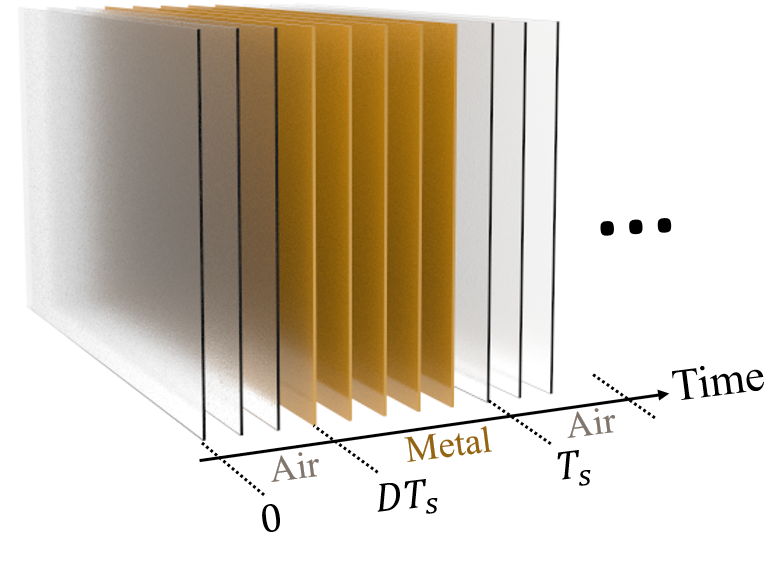}}
        \hspace{0.1cm}
        \subfloat[]{\includegraphics[width=0.22\textwidth]{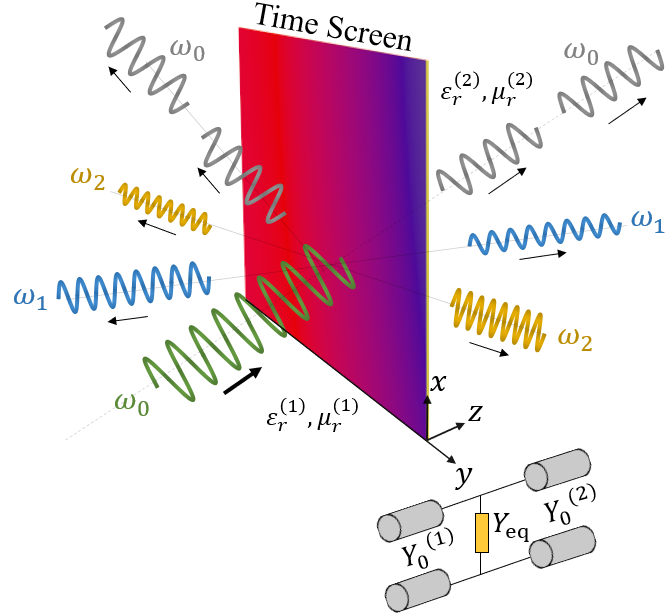}}
        \caption{(a) Time evolution of the proposed configuration. The screen periodically alternates between ``air" and ``metal" states. (b) Illustration of the spacetime diffraction caused by the time-periodic screen and its equivalent circuit. }
        \label{fig1}
\end{figure}   

Transmission-line and ABCD-parameter models have already been employed in electromagnetic systems with instantaneous temporal interfaces \cite{Xiao2014, Elnaggar2020}. Such are the cases reported in \cite{ramaccia2020, Ramaccia2021} and more recently in \cite{Stefanini2022}. The equivalent circuit aids for a better understanding of the situations there described. However, in most cases, no periodic modulation exists and there is no excitation of higher-order harmonics. The work in \cite{Alex2022TV} considers a system formed by a metallic screen suffering a periodic modulation. The system is fed by an external plane wave, exciting an infinite number of periodic Floquet harmonics. The paper reports the derivation of the circuit model but no many situations are evaluated. The present work is intended to exploit the model possibilities, increasing the number of modulation ratios, introducing the concepts of \emph{macroperiod} and \emph{duty cycle} to the time modulation, with the objective of enriching the diffraction phenomenon. In addition, the scattering parameters are quantitatively evaluated, constituting a novelty with respect previous works in the literature.    


The structure under consideration is sketched in Fig.~\ref{fig1}. A monochromatic plane wave of frequency $\omega_0$ illuminates a time metamaterial that periodically alternates between ``air" and ``metal" (perfect electric conductor, PEC) states, as represented in Fig.~\ref{fig1}(a). The time screen is considered to be infinitesimally thin along the $z$ axis and very large in $x$ and $y$ directions [see Fig.~\ref{fig1}(b)].  The time periodicity of the varying screen is $T_{\text{s}} = 2\pi/\omega_{\text{s}}$, from which the whole cycle repeats. In the more general scenario, the time screen could remain in air state ($DT_{\text{s}}$) for a different time than it remains in metal state ($[1-D]T_{\text{s}}$). Here, $D \in [0, 1]$ is the \emph{duty cycle} of the time modulation. Extreme cases $D=0$ and $D=1$ would imply that the time screen remains invariant in metal and air states the whole time, respectively. The fact of varying the duty cycle $D$ and its implications were not discussed in our previous work \cite{Alex2022TV}, since a fixed value of $D=0.5$ was implicitly assumed. As it will be detailed later, modifying the duty cycle enriches the diffraction phenomenon, since half-period temporal symmetry is broken and this leads to asymmetries in harmonic excitation. The \emph{modulation ratio} $F = \omega_0 / \omega_{\text{s}} = T_{\text{s}} / T_{0}$ constitutes a second factor to be discussed. The nature of the reflected and transmitted fields across the discontinuity directly depends on this parameter, and as it will be discussed below, it may govern the power transfer between different harmonics.

Ideally, the implementation of a time-periodic thin screen that transits between "metal" and "air" states would require of a reconfigurable material  whose electrical properties can be tuned in real time. Two-dimensional materials such as graphene and its oxides \cite{Yu2020}, or monolayer molybnedum disulfide (MOS$_2$) \cite{LI201533} and  hexagonal boron nitride (h-BN) \cite{Laturia2018} are potential candidates for this purpose. For instance, it is well known that biased graphene can behave as a good electrical conductor (low surface-resistance value),  being able to recreate the metal state. As the bias conditions are relaxed, the surface resistance of graphene increases, leading to absorption and transparent ("air"-like) states \cite{Allen2010, Zhu_2014}. This properties of graphene have been exploited for the design of reconfigurable devices with advanced functionalities \cite{Molero2021}. 

The guidelines to derive the equivalent circuit are elaborately reported in \cite{Alex2022TV}. The propagation of the incident and reflected waves, and the transmitted one are represented by transmission lines with $Y_{0}^{(1)}$ and $Y_{0}^{(2)}$ characteristic admittances, respectively. The equivalent admittance $Y_{\text{eq}}$ accounts for the effect of the \emph{time discontinuity}, and includes the effect of all the higher-order harmonics $E_n$. In general, Floquet coefficients $E_n$ are computed as
\begin{align}
    \label{ordenn} E_{n}^{(1)} &= E_{n}^{(2)} = \frac{1}{T_{\text{m}}} \displaystyle\int_{0} ^{T_{\text{m}}} E(t) \text{e}^{-\text{j} \omega_{n} t} \text{d}t\,,
\end{align}
where $\omega_n = \omega_0 + n2\pi/T_{\text{m}}$ is the angular frequency associated to the $n$-th Floquet harmonic. Moreover, the reflection ($R$) and transmission ($T$) coefficients, both associated to the fundamental harmonic ($n=0$), can be directly estimated from the circuit model as
\begin{equation}\label{R2}
    R = \frac{Y_{0}^{(1)} - Y_{0}^{(2)} -  Y_\mathrm{eq}}{Y_{0}^{(1)} + Y_{0}^{(2)} +  Y_\mathrm{eq}},
\end{equation}
\begin{equation}
    T = 1 + R\, .
\end{equation}
\begin{figure*}[!t]
	\centering
        \subfloat[]{\includegraphics[width=0.33\textwidth]{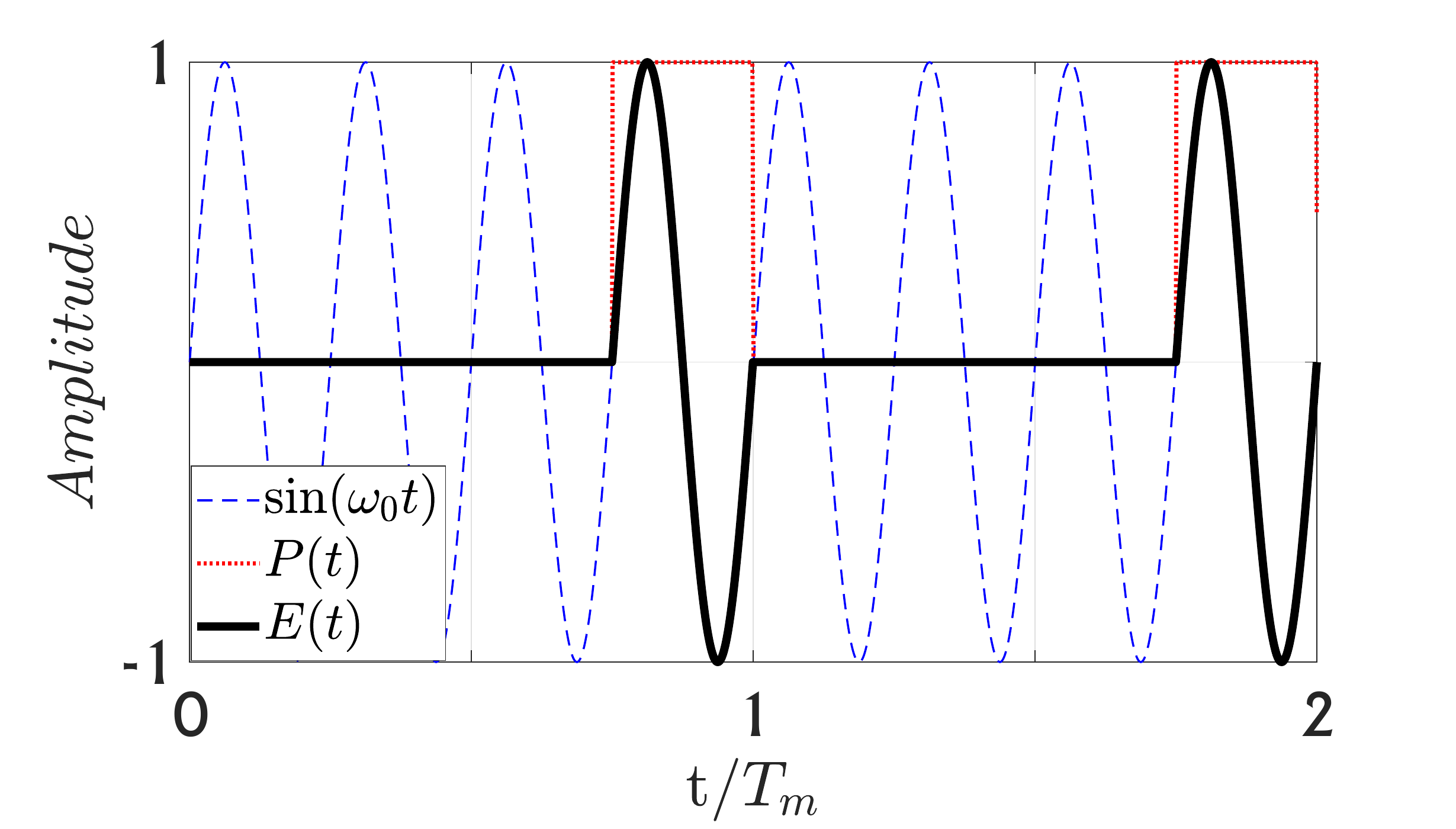}}
        \subfloat[]{\includegraphics[width=0.33\textwidth]{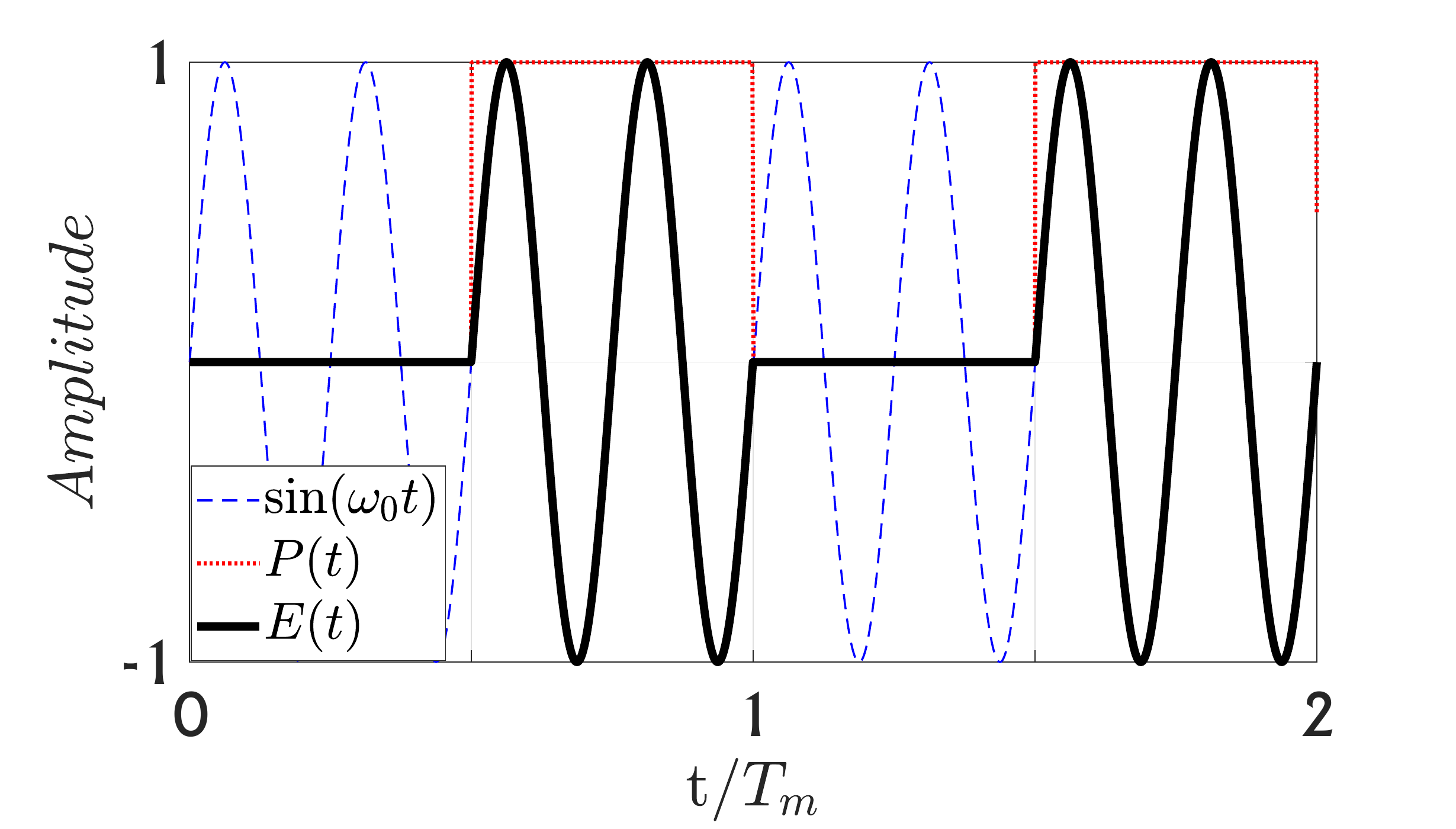}}
        \subfloat[]{\includegraphics[width=0.33\textwidth]{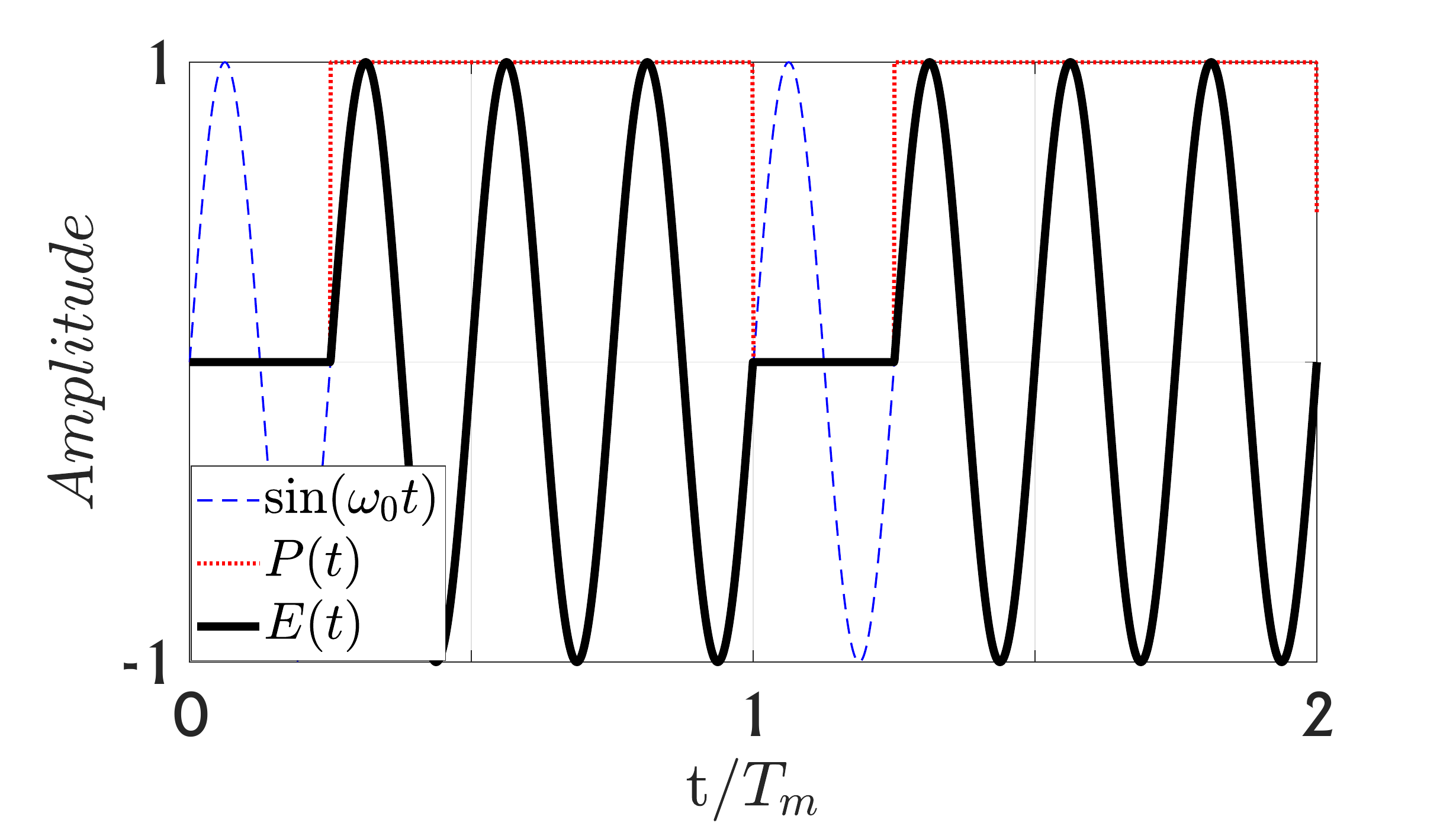}}
        
        \subfloat[]{\includegraphics[width=0.33\textwidth]{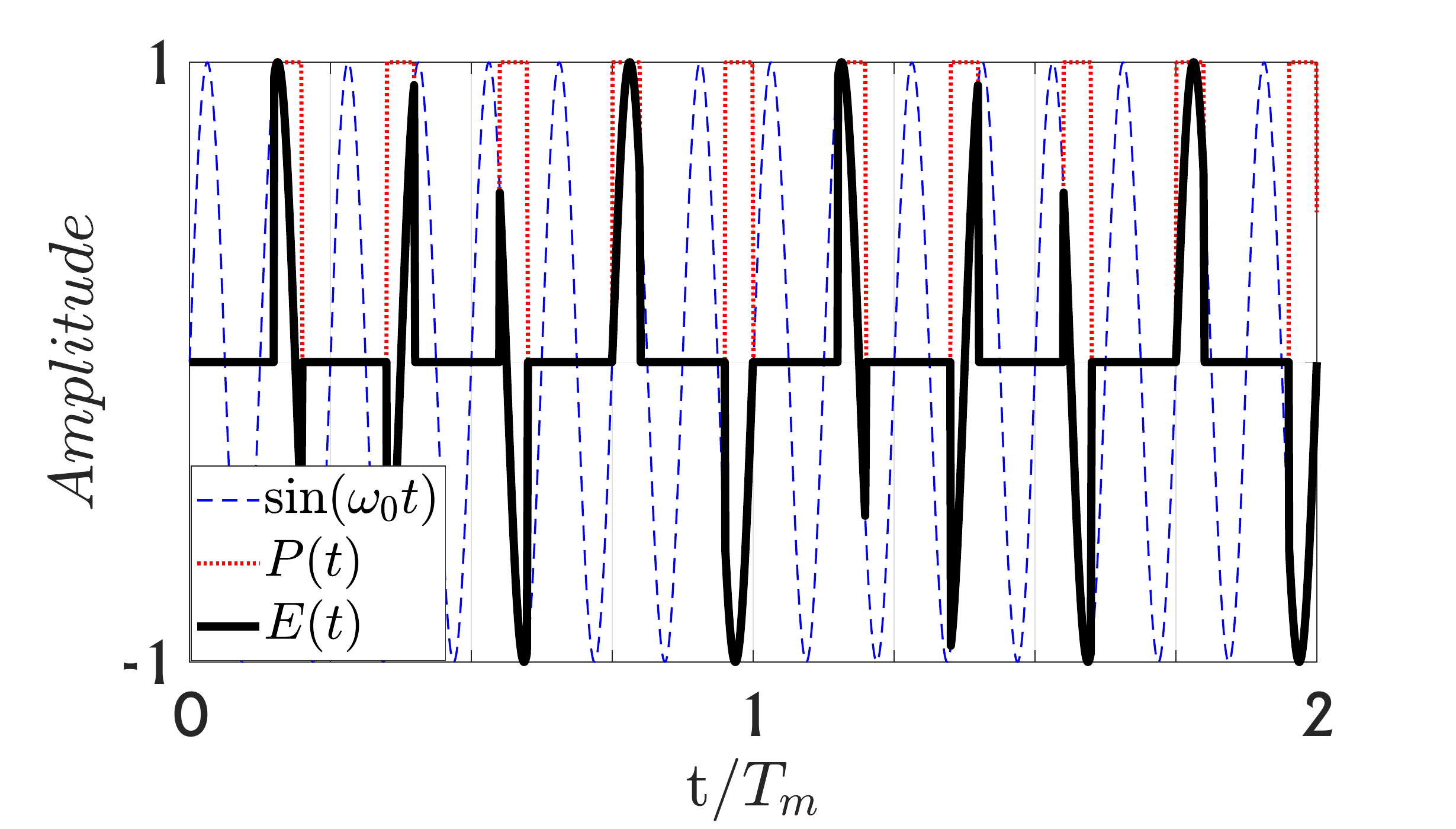}}
        \subfloat[]{\includegraphics[width=0.33\textwidth]{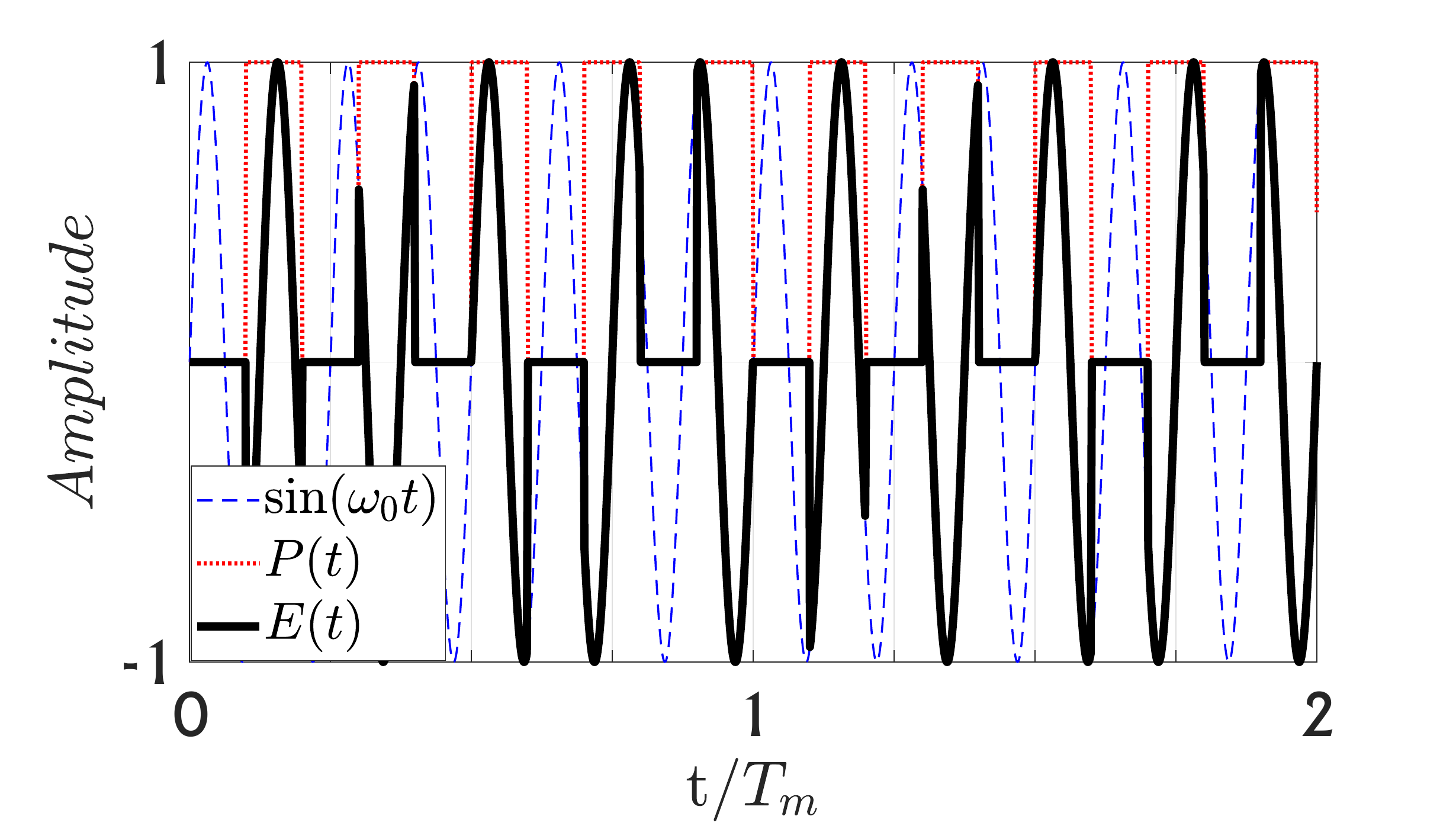}}
        \subfloat[]{\includegraphics[width=0.33\textwidth]{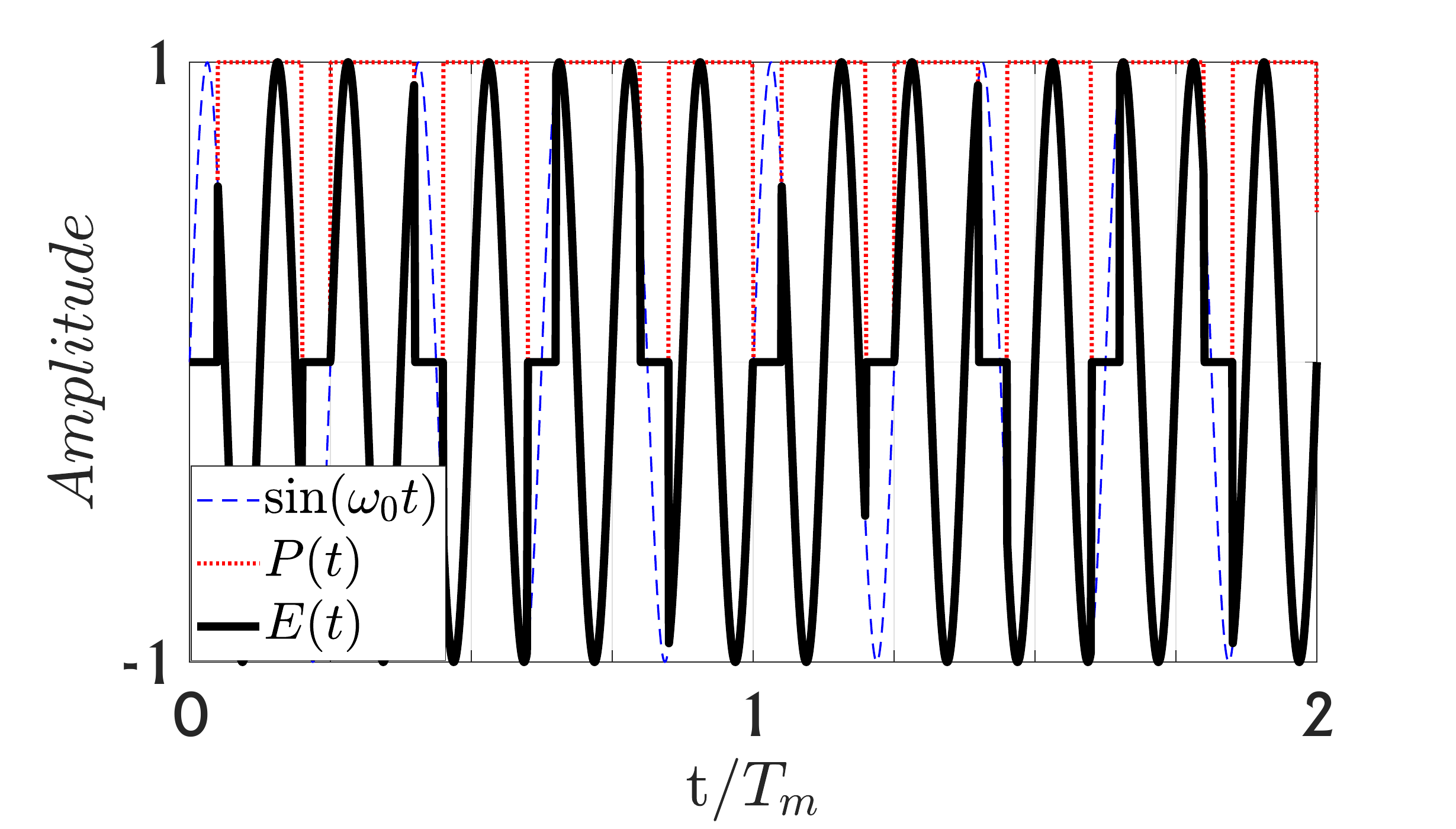}}
	\caption{\small Two macroperiods of $E(t)$ when: (a) $F=4$, $D=0.25$, (b) $F=4$, $D=0.5$, (c) $F=4$, $D=0.75$, (d) $F=1.6$,  $D=0.25$, (e) $F=1.6$, $D=0.5$, (f) $F=1.6$, $D=0.75$.} 
	\label{PERIODOS}
\end{figure*}

The coupling between harmonics, described in terms of transformers with turn ratio $N(\omega_{n})$, demands a previous knowledge of the field profile $\mathbf{E}(t)$ at the discontinuity along a time period. Our previous work is focused on integer time-modulation ratios $F$, assuming $\omega_{\text{s}} \le \omega_{0}$ in most cases. This is a very restricted situation. The extension from integer to rational (not irrational) modulation ratios is here taken into account, modifying the way to get $\mathbf{E}(t)$. Now, $\mathbf{E}(t)$ is influenced by $D$ and $F$, leading to the definition of the term \emph{macroperiod}. A macroperiod $T_{\text{m}}$ is defined as the minimum time periodicity where both the incident-wave vibration ($\omega_{0}$) and the screen variation ($\omega_{\text{s}}$) complete a full cycle simultaneously. 

Mathematically, every \emph{rational} modulation ratio $F$ can be approximated by a fraction of two integers, $F_N$ and $F_D$, according to $F = F_N / F_D$. Since $F$ was previously defined as $F=T_\mathrm{s}/T_0$, the temporal macroperiod $T_{\text{m}}$ must follow the condition
\begin{equation}
    T_{\text{m}} = F_N T_0 = F_D T_\mathrm{s}\,.
\end{equation}

\begin{figure*}[!t]
	\centering
        \subfloat[]{\includegraphics[width=0.33\textwidth]{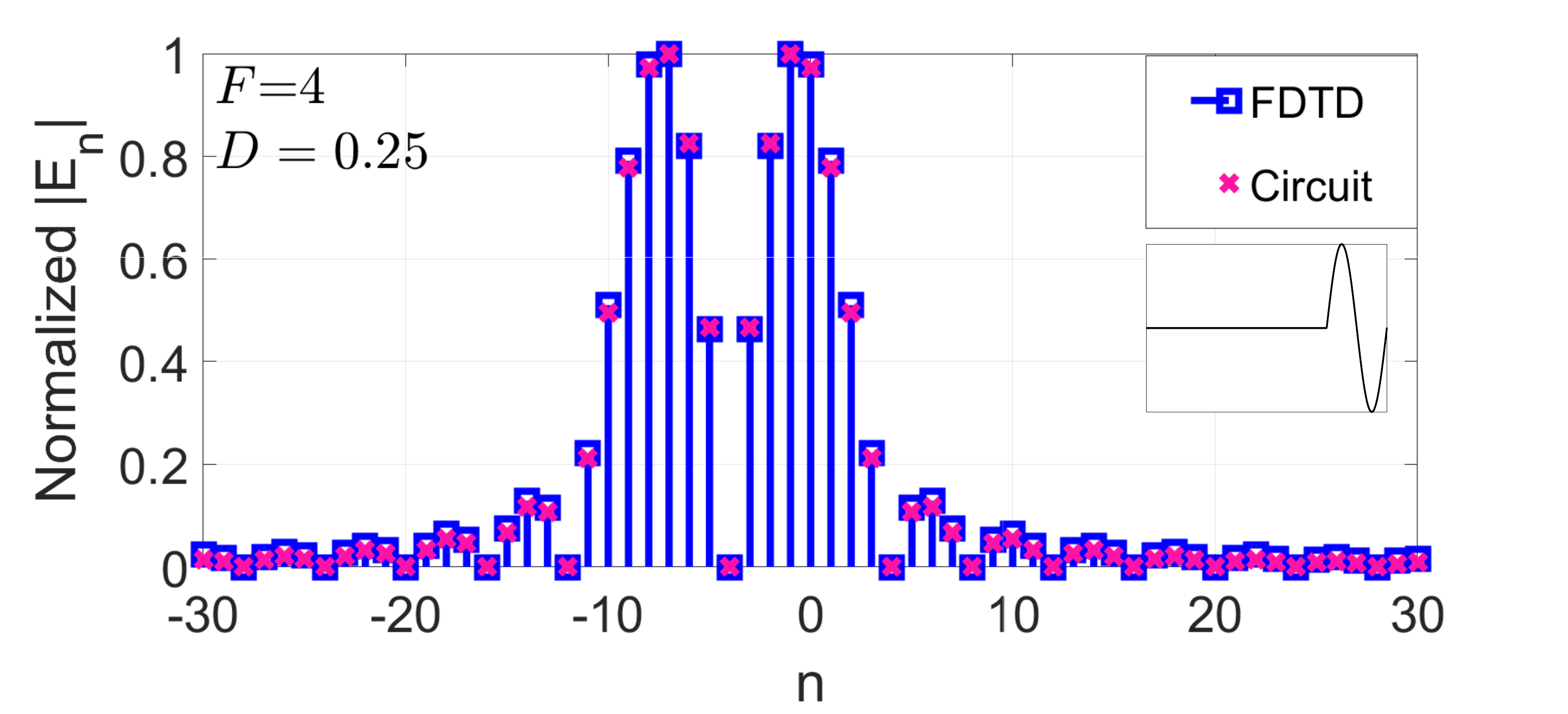}}
        \subfloat[]{\includegraphics[width=0.33\textwidth]{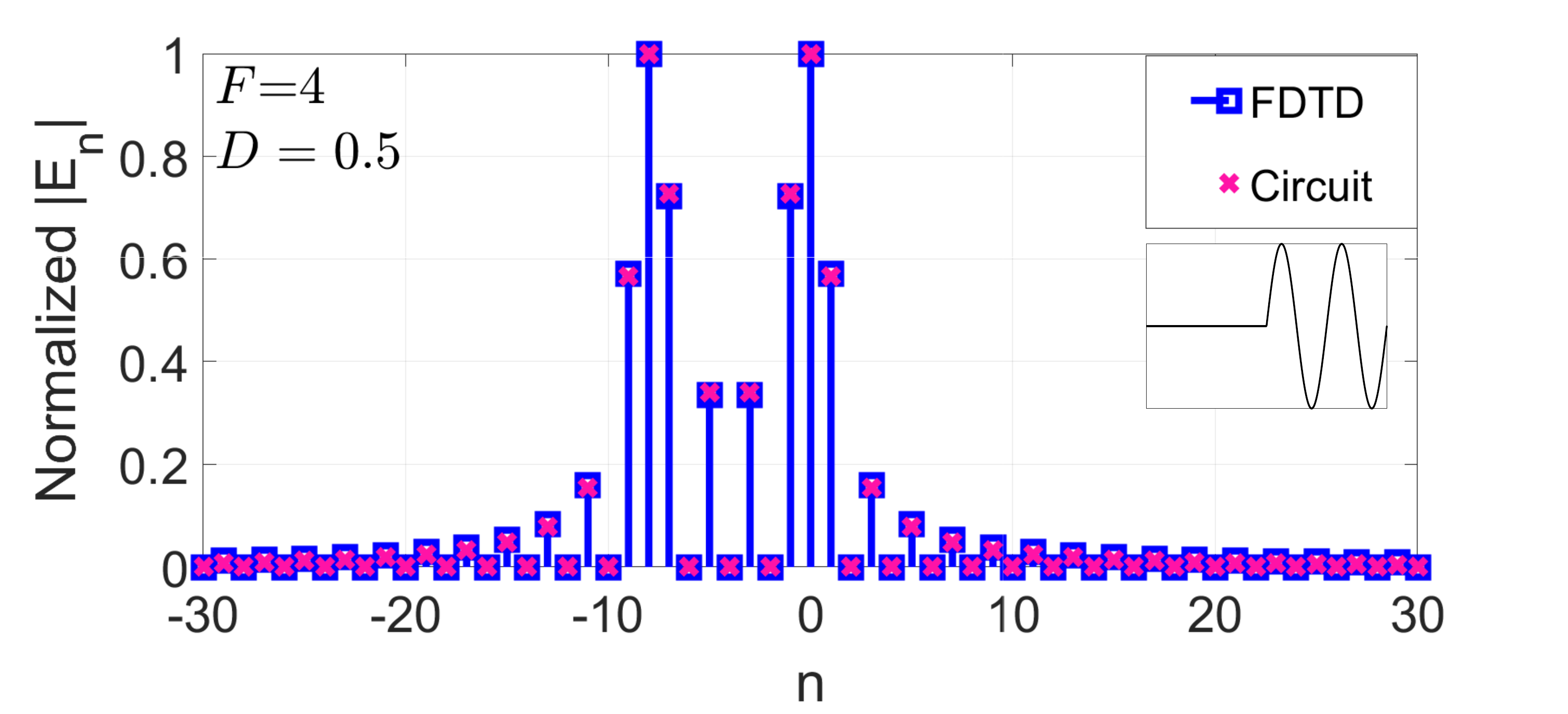}}	
        \subfloat[]{\includegraphics[width=0.33\textwidth]{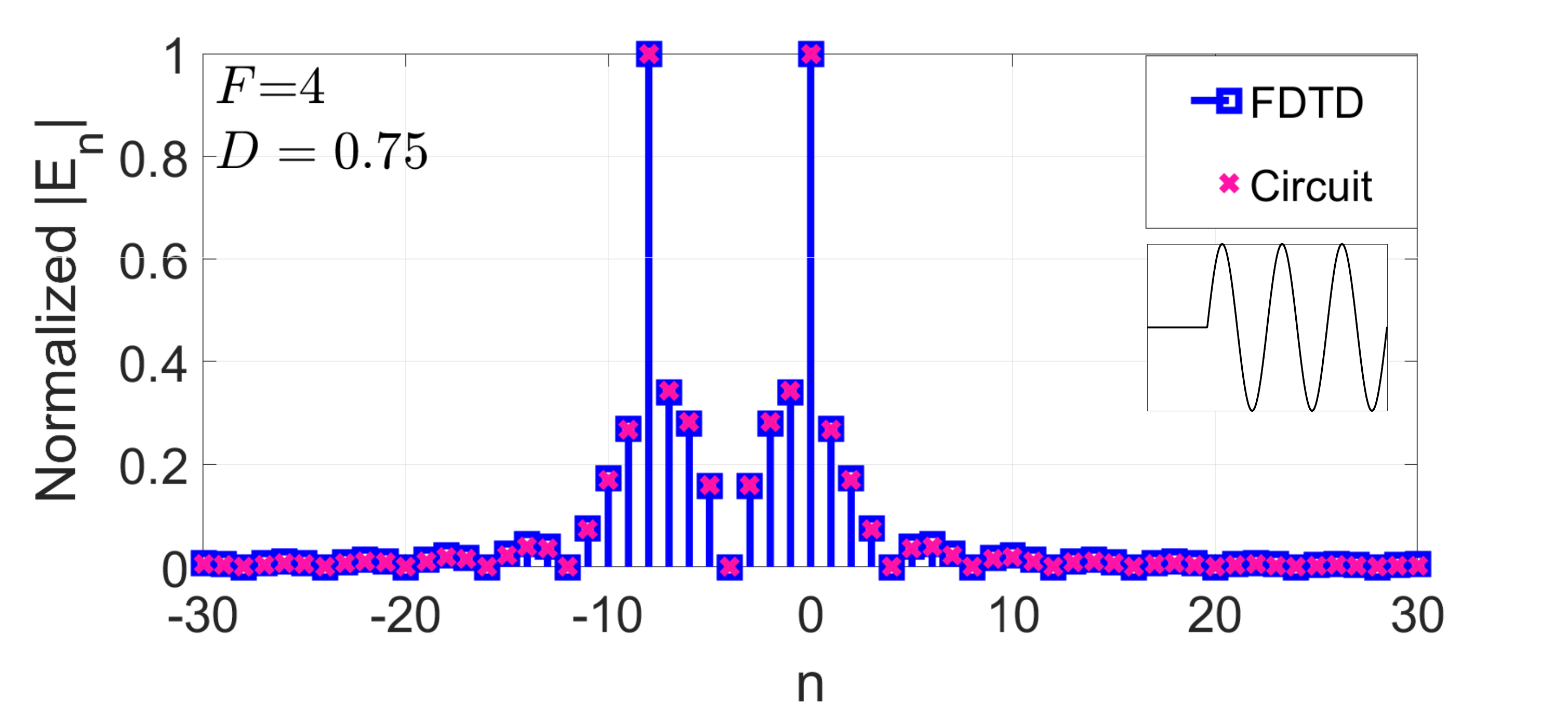}}
 
        \subfloat[]{\includegraphics[width=0.33\textwidth]{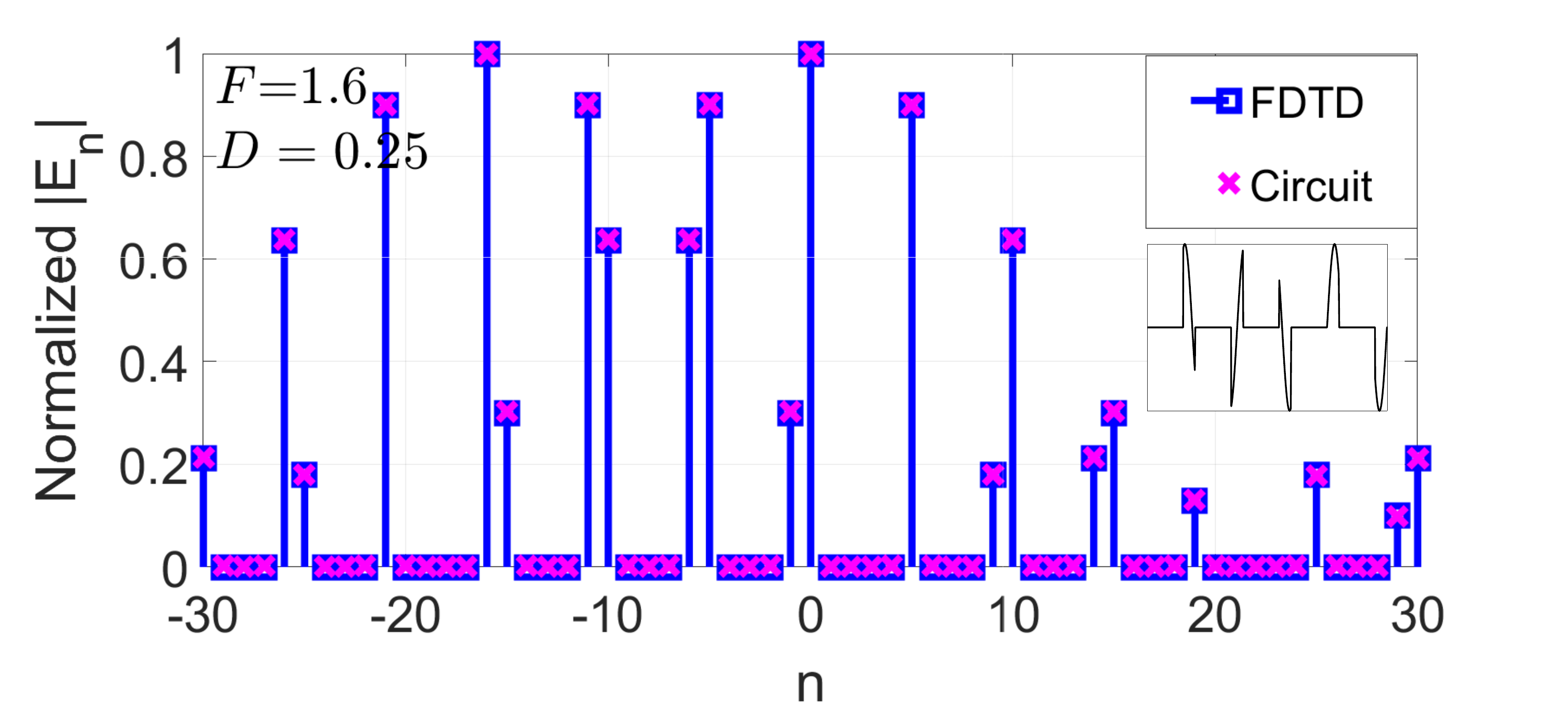}}
        \subfloat[]{\includegraphics[width=0.33\textwidth]{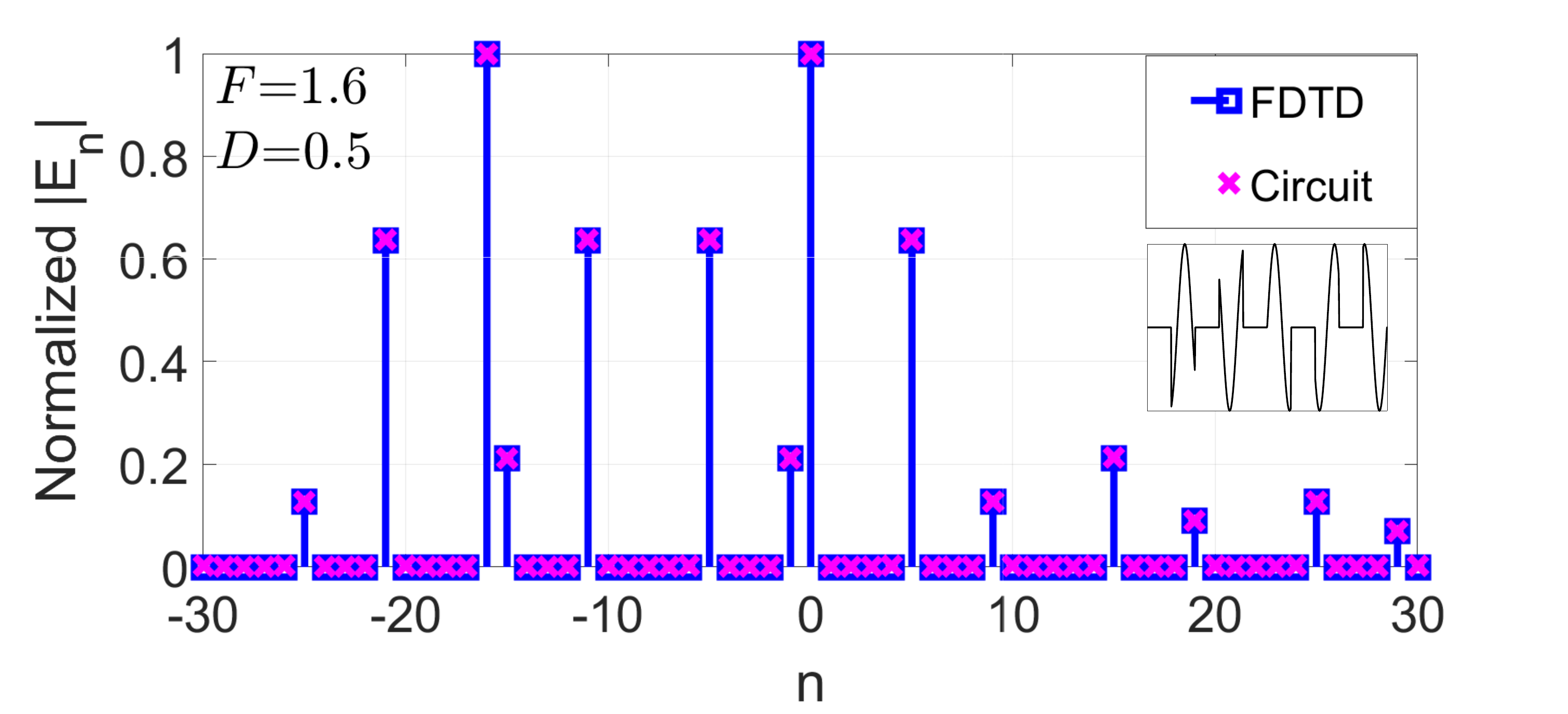}}
        \subfloat[]{\includegraphics[width=0.33\textwidth]{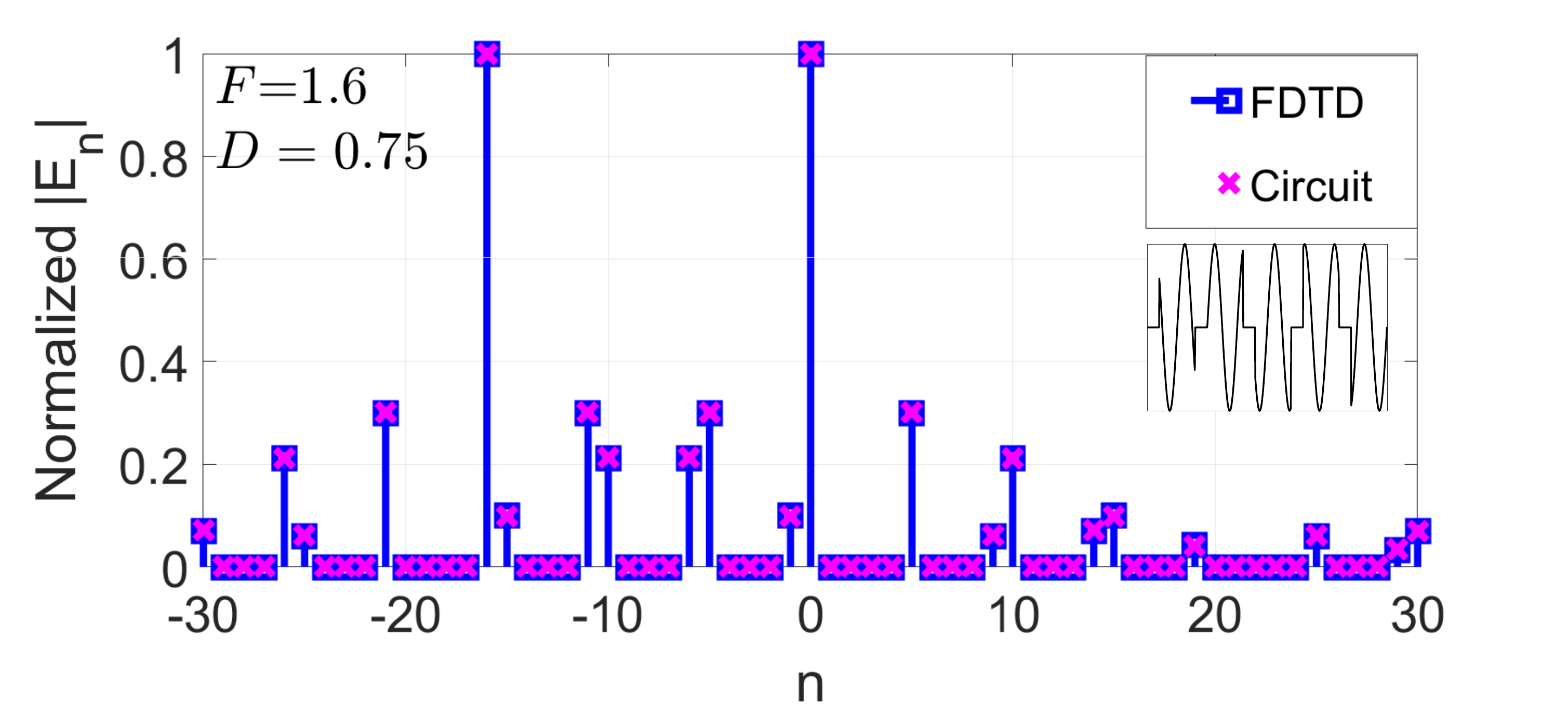}}
  	\caption{\small Normalized amplitude of the Floquet coefficients in the cases: (a) $F=4$, $D=0.25$, (b) $F=4$, $D=0.5$, (c) $F=4$, $D=0.75$, (d) $F=1.6$,  $D=0.25$, (e) $F=1.6$, $D=0.5$, (f) $F=1.6$, $D=0.75$. Normal incidence is assumed.} 
	\label{Nomalcoeficients}
\end{figure*}
Thus, a macroperiod is completed after $F_N$ and $F_D$ cycles for the incident wave ($T_0$) and the time modulation ($T_s$), respectively. Please note that an \emph{irrational} modulation ratio $F$ cannot be described in terms of a fraction of two integers, leading to an infinite set of decimals. As a consequence, the macroperiod of an irrational modulation ratio would be infinite and the formulation proposed here would not be applicable since time periodicity is lost. Thus, the field profile $\mathbf{E}(t)$ is therefore defined along a macroperiod, ensuring a stationary situation. It can be mathematically described as
\begin{equation}
    \mathbf{E}(t) = \sin(\omega_0 \,t) \, P(t) \,\hat{\mathbf{y}}, \quad t\in [0, T_{\text{m}}],
\end{equation}
where $P(t)$ is a pulse train of period $T_{\text{s}}$ and duty cycle $D$.


\Figs{PERIODOS}(a)-(c) depict the evolution of $E(t) = |\mathbf{E}(t)|$, when the modulation ratio is fixed to $F=4$ ($F=F_N/F_D = 4/1$), for duty cycles $D = 0.25, 0.5, 0.75$, respectively. In these cases, the value of the $macroperiod$ $T_{\text{m}}$ coincides with $T_{\text{s}}$ (or $4T_{0}$). A second case regarding $F$ as a rational number is exhibited in \Figs{PERIODOS}(d)-(f), where it can be appreciated how the shape of $E(t)$ becomes more complex. Now $F=1.6 = 8/5$, increasing the macroperiod up to $T_{\text{m}} = 5T_{\text{s}}$ or, analogously, $T_{\text{m}} = 8T_{0}$. In all these figures $E(t)$ is drawn in a time interval defined by two consecutive macroperiods, in order to appreciate the existing periodicity. As will be explained below, the variation of $D$ has direct implications on the amplitude provided by each Floquet harmonic.

A correct definition of $E(t)$ is crucial to guarantee accurate predictions by the circuit model. A first test of the validity of the circuit approach is shown in \Fig{Nomalcoeficients}. It illustrates the normalized spectral response of the transmitted field in the cases reported in \Fig{PERIODOS}, with an inset showing the field profile $E(t)$. A TM-polarized plane wave impinging normally has been assumed for the computation. As expected, the spectrum is split in discrete harmonics, whose amplitudes vary for each case.  Together with the results provided by the equivalent circuit, numerical results extracted by  self-implemented finite-different time-domain (FDTD) are included. FDTD methods \cite{Stewart2018, Vahabzadeh2018} have proven to be interesting numerical alternatives  to validate analytical results due to the absence of specific commercial electromagnetic solvers oriented to deal with spacetime metamaterials. It is also worthy to emphasise that due to assumption of normal incidence, all the harmonics are propagative (there is no harmonics with evanescent nature) and moreover, they leave the air-metal interface at the incidence direction ($\theta_{n}=0^{\text{o}}$). This result comes from Eq.[28] in \cite{Alex2022TV}
\begin{equation}\label{angle}
    \theta_n^{(i)} = \arctan\left(\dfrac{k_{t}}{\sqrt{\varepsilon_r^{(i)} \mu_r^{(i)}\left[\frac{\omega_0 +  2\pi n / T_\mathrm{m}}{c}\right]^2 - k_{t}^2}}\right)\,
\end{equation}
when imposing $k_{\text{t}} = 0$ with $k_{\text{t}}$ being the transverse wavevector of the incident wave, and $i = 1, 2$ being the index indicating the leftmost/ rightmost medium respectively.

As visualized in \Figs{Nomalcoeficients}(a)-(c) for $F = 4$, the value of $D$  modifies the amplitude of the harmonics. For instance, when the wave encounters free-space in a longer time interval than metal at the interface, $D=0.75$, the biggest amplitude values are carried by the fundamental harmonic ($n=0$) and that with order $n=-8$ [see \Fig{Nomalcoeficients}(c)]. When this interval decreases to $D=0.25$, the amplitude of these predominant harmonics reduces with respect the rest of diffracted harmonics [see \Fig{Nomalcoeficients}(a)]. This tendency seems to be progressive if we check \Figs{Nomalcoeficients}(a)-(c) from left to right. If the modulation ratio is varied down to $F = 1.6$, as illustrated in the spectra shown in \Figs{Nomalcoeficients}(d)-(f), it can be noticed that the modal distance between harmonics have changed. This fact can be appreciated since those carrying  more energy are now the fundamental one ($n = 0$) and the one with order $n=-16$. In general, increasing the duty cycle $D$ provokes that the time screen remains in "air" state a greater amount of time. Thus, the field profile $\mathbf{E}(t)$ progressively turns into the original incident plane wave as $D$ approaches the unit. Therefore, the spectrum of the system resembles the spectrum of a conventional sine function, predominated by two delta functions at frequencies $\pm \omega_0$, with the rest of harmonics being significantly attenuated. This phenomenon is observed in \Figs{Nomalcoeficients}(a)-(c) and \Figs{Nomalcoeficients}(d)-(f) as $D$ is increased. Note that breaking the perfect temporal symmetry of the modulation ($D\neq 0.5$) causes that harmonics of even and odd nature excite indistinctly. The situation was different in our previous work \cite{Alex2022TV}, where the duty cycle was fixed to $D=0.5$. In that case, perfect temporal symmetry provoked that higher-order even harmonics became null, fact that can be also appreciated in  \Figs{Nomalcoeficients}(b) and (e). Therefore, the introduction of the duty cycle to the time modulation enriches the diffraction spectrum, which is of potential interest for the development of time-based beamformers.

\begin{figure}[t!]
	\centering
        \subfloat[]{\includegraphics[width=0.4\textwidth]{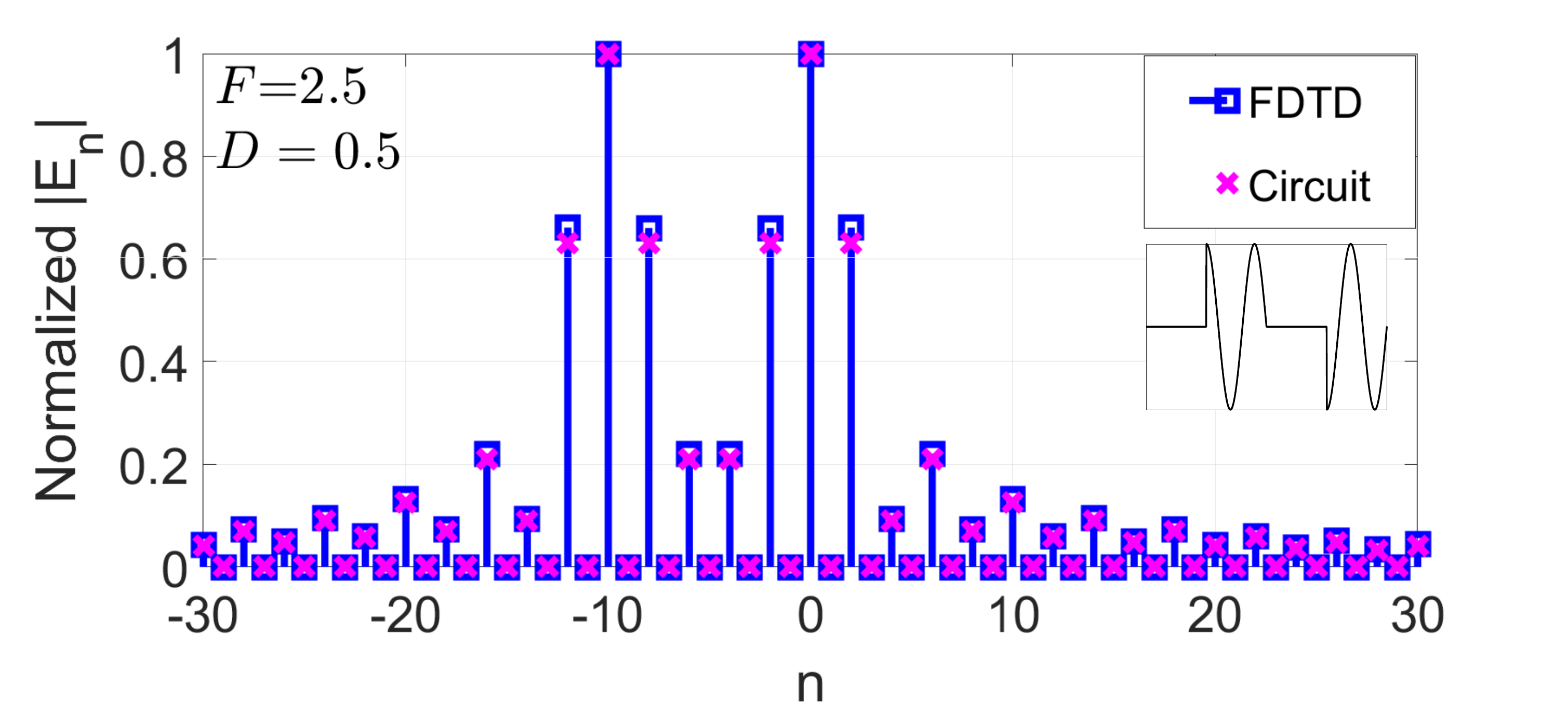}}
	\hspace{0.01in}	
        \subfloat[]{\includegraphics[width=0.4\textwidth]{N_D_0p5_f_1p6_FINAL.pdf}}
	\hspace{0.01in}	
        \subfloat[]{\includegraphics[width=0.4\textwidth]{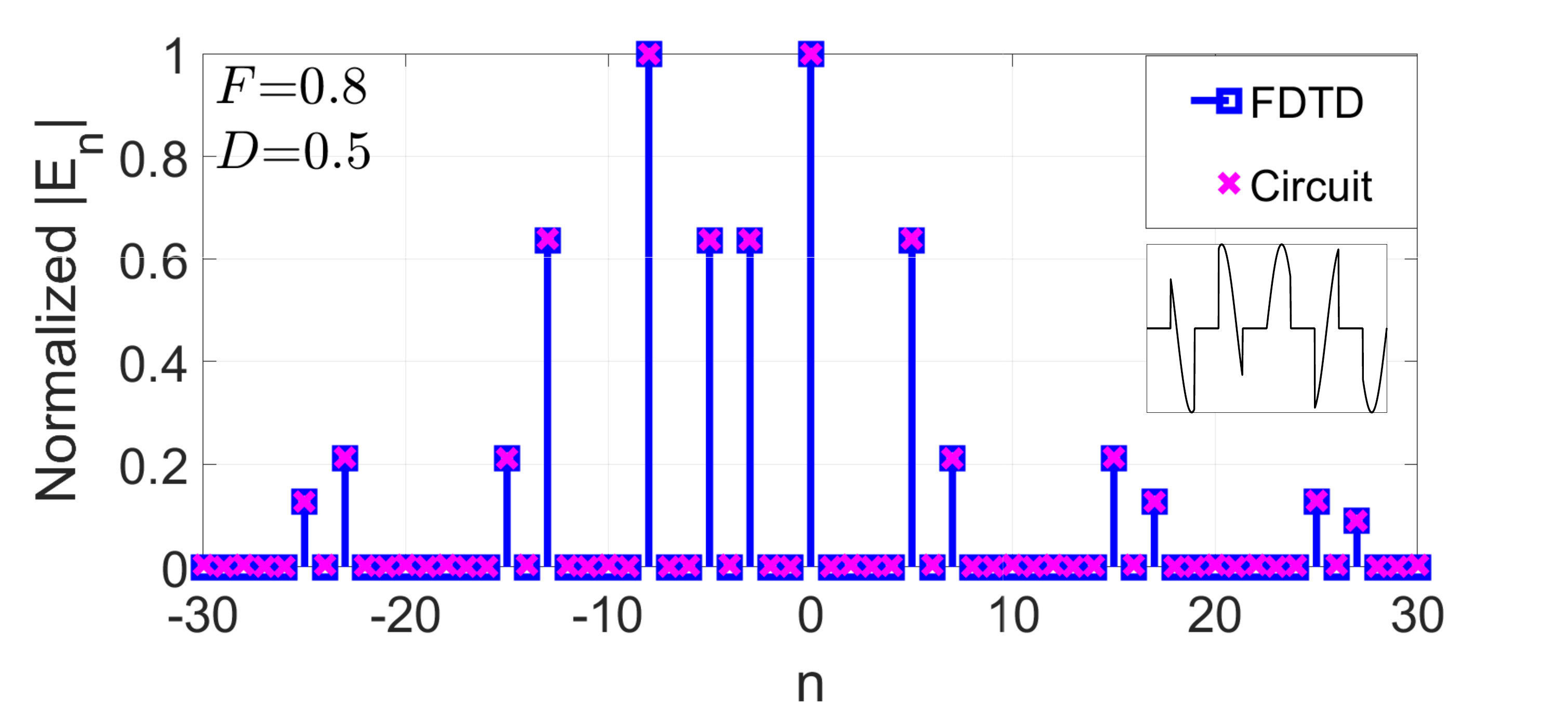}}
	\hspace{0.01in}	
	\caption{\small Normalized amplitude of the Floquet coefficients in the cases: (a) $F=2.5$, $D=0.5$, (b) $F=1.6$, $D=0.5$, (c) $F=0.8$, $D=0.5$. Oblique incidence is assumed: $\theta_{\text{inc}}=30^{\text{o}}$.} 
	\label{Fig4coeficientes}
\end{figure}

To understand the effect of the reconfigurability in this time-periodic metamaterial, \Fig{Fig4coeficientes} shows configurations with different modulation ratios $F$ while keeping the same duty cycle fixed to $D=0.5$. This situation consisting is well captured by the circuit model, after a previous definition of $E(t)$. The temporal evolution of $E(t)$ along a macroperiod is included as an inset of the figures. Now, TE oblique incidence is assumed under an angle of incidence $\theta_{\text{inc}}=30^{\text{o}}$. The transverse wavevector is no longer null ($k_{\text{t}} \ne 0$), opening the possibility to excite evanescent harmonics according to \eqref{angle}. \Fig{Fig4coeficientes}(a) depicts a first case governed by $F = 2.5$. For this configuration, some evanescent harmonics have non-zero amplitude values, as those with orders $n=-6, -4$. The rest of harmonics with non-zero amplitude are propagative. As $F$ changes, the amplitude distribution get modified. In case illustrated in \Fig{Fig4coeficientes}(b) the modulation ratio is $F=1.6$, and now the evanescent harmonics with significant amplitude are those with orders $n=-11, -5$. For $F = 0.8$, reported in \Fig{Fig4coeficientes}(c), they become the ones with orders $n=-5, -3$.

\begin{table}[!t]
\caption{Diffraction angle $\theta_n$ of the main Floquet harmonics while applying different modulation ratios. Oblique incidence is considered: $\theta_{\text{inc}}=30^{\text{o}}$. The duty cycle of the time-periodic screen is $D=0.5$.}
\centering \label{table1}
\begin{tabular}{c *5c} 
\toprule
 & \multicolumn{1}{c}{\textbf{Diffracted Angle}} & \textbf{Circuit}  & \textbf{FDTD}    \\ 
\midrule
\multirow{3}{1.5cm}{$\boldsymbol{F=2.5}$} & $\boldsymbol{\theta_{-2}} \boldsymbol{(^\mathrm{o})}$ & $56.44$ & $56.49$  \\ 
 & $\boldsymbol{\theta_0} \boldsymbol{(^\mathrm{o})}$  &  $30$ &  $29.97$ \\
 & $\boldsymbol{\theta_2} \boldsymbol{(^\mathrm{o})}$  &  $20$ &  $20.05$ \\
\midrule
\multirow{2}{1.5cm}{$\boldsymbol{F=1.6}$}  & $\boldsymbol{\theta_0} \boldsymbol{(^\mathrm{o})}$ & $30$ & $30.05$  \\ 
 & $\boldsymbol{\theta_5} \boldsymbol{(^\mathrm{o})}$ &  $17.92$ &  $17.92$ \\
\midrule
\multirow{2}{1.5cm}{$\boldsymbol{F=0.8}$}  & $\boldsymbol{\theta_0} \boldsymbol{(^\mathrm{o})}$ & $30$ & $30.01$   \\ 
 & $\boldsymbol{\theta_5} \boldsymbol{(^\mathrm{o})}$ &  $12.89$ &  $12.91$   \\
 \bottomrule
 \end{tabular}
\end{table}
\begin{figure}[!t]
	\centering
        \subfloat[]{\includegraphics[width=0.4\textwidth]{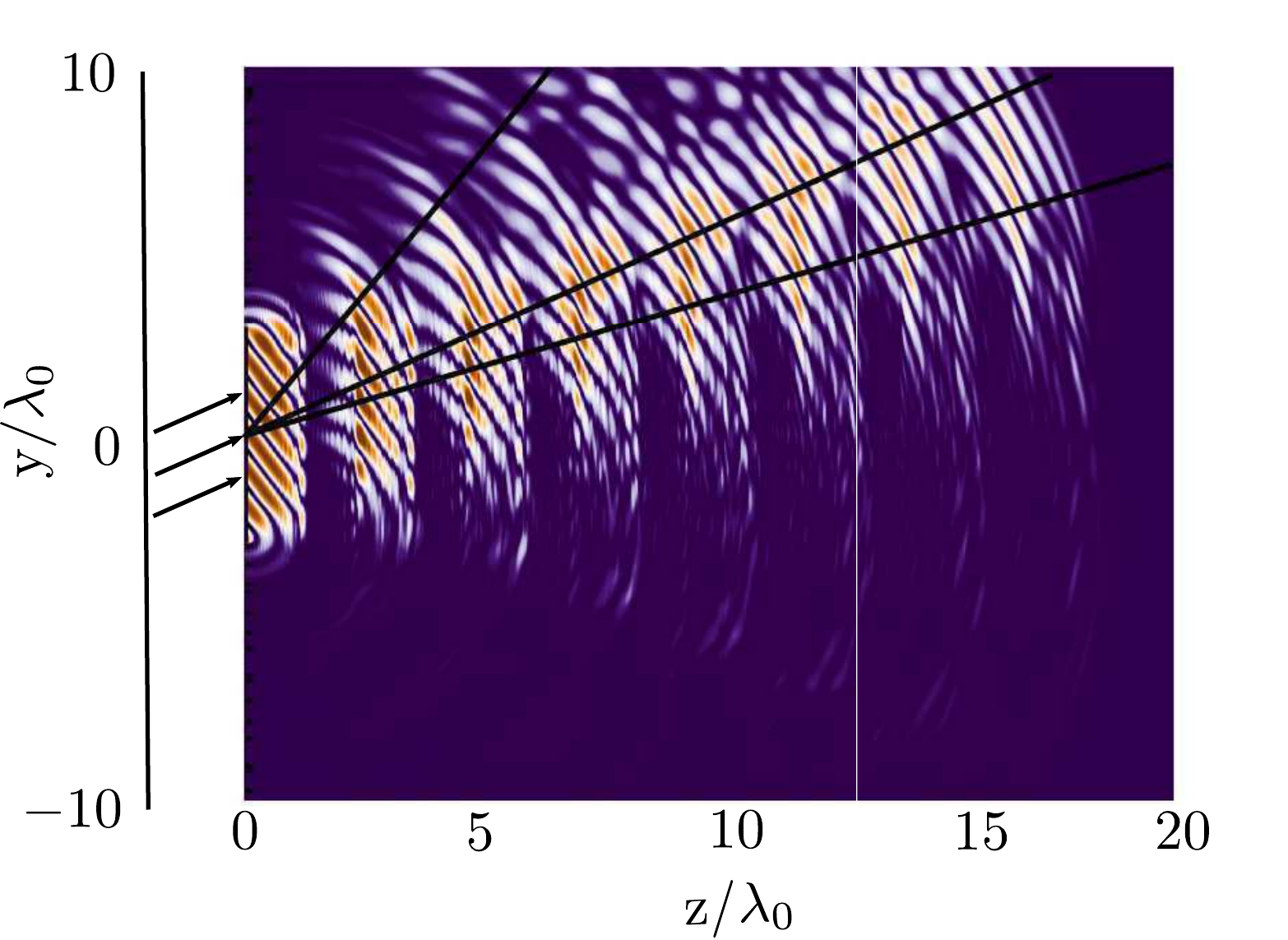}}
	\hspace{0.01in} 
        \subfloat[]{\includegraphics[width=0.4\textwidth]{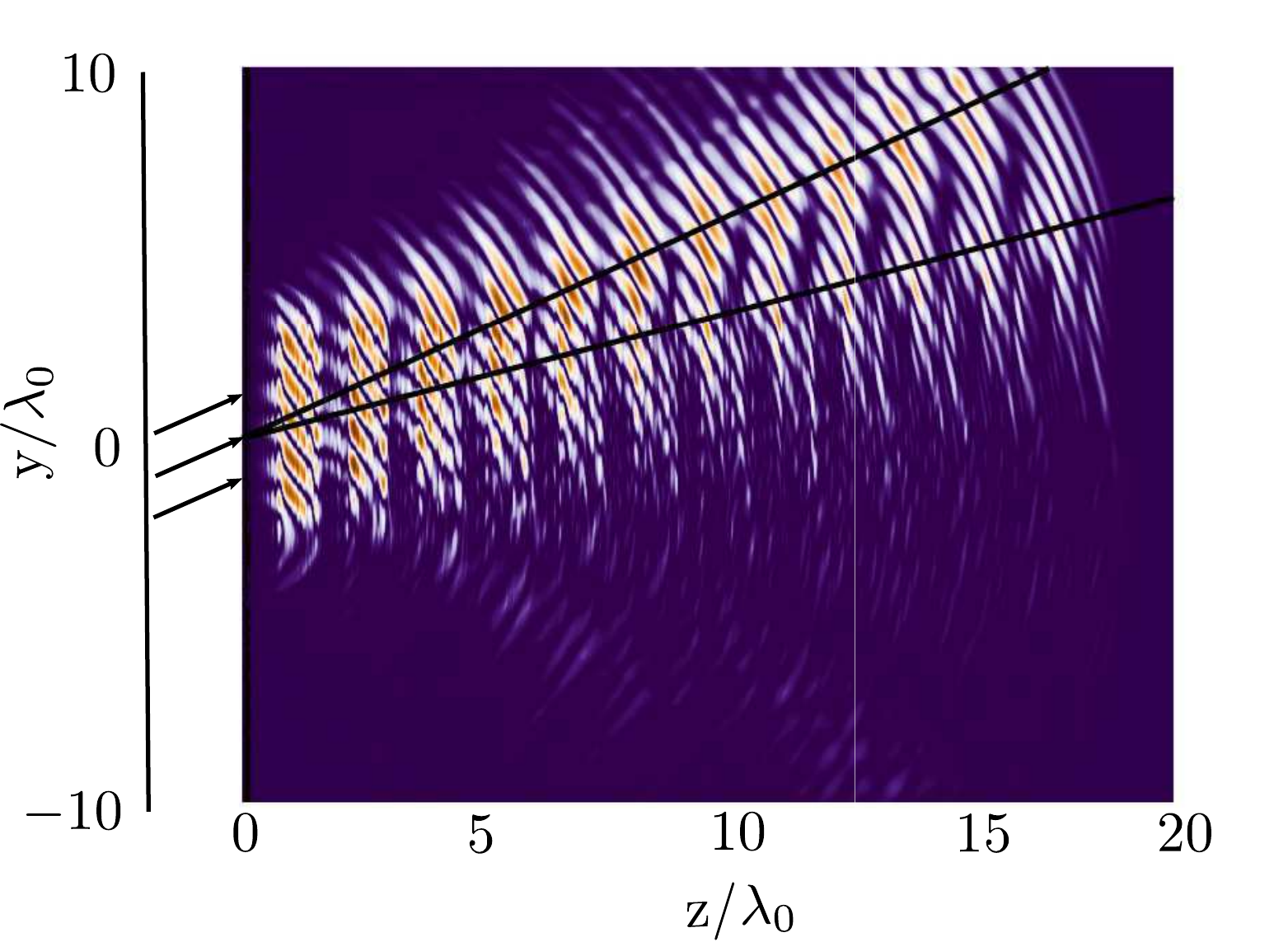}}
	\hspace{0.01in}	
        \subfloat[]{\includegraphics[width=0.4\textwidth]{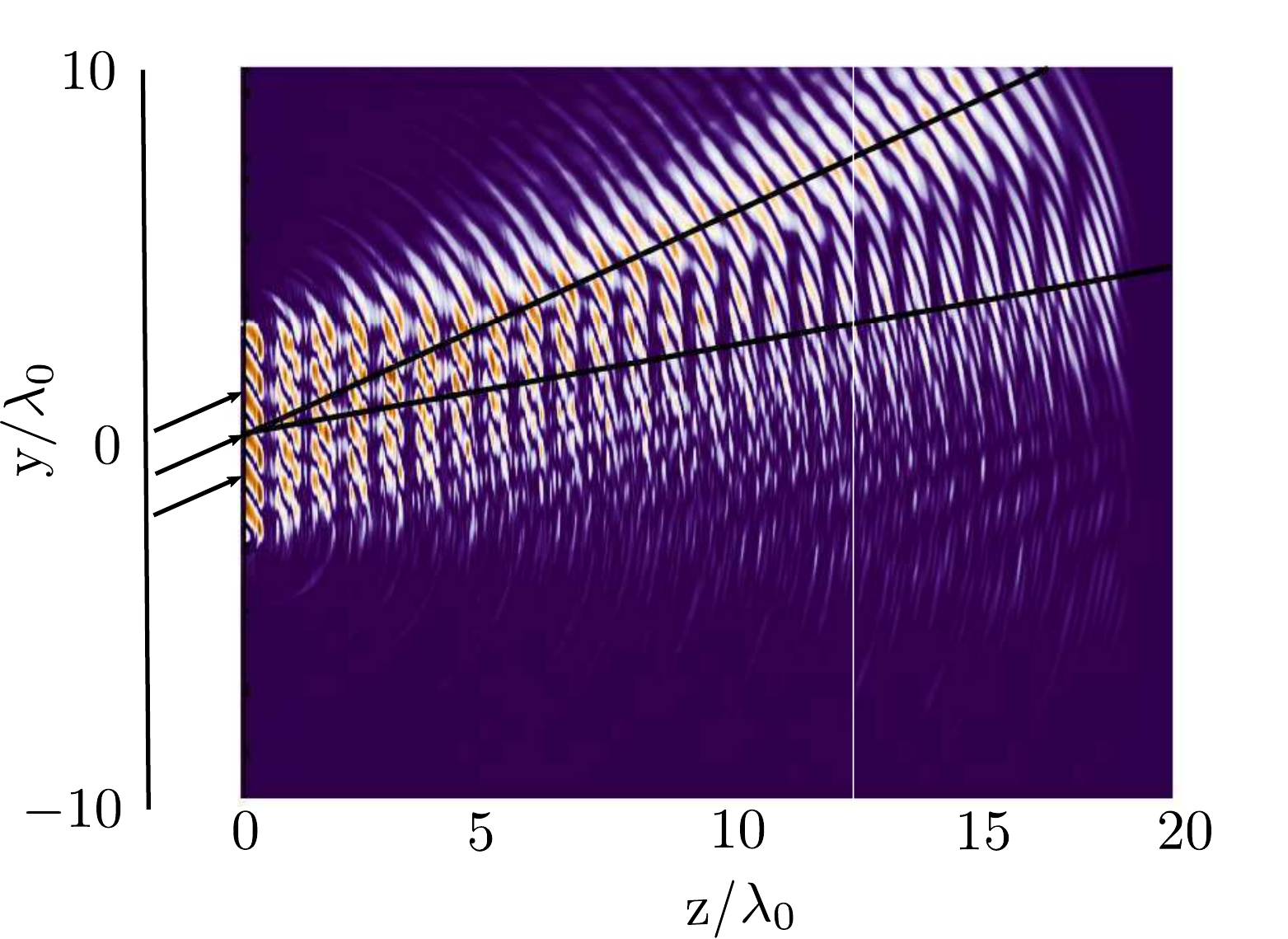}}
	\hspace{0.01in}	
	\caption{\small Electric field distribution obtained with the FDTD method in the cases: (a) $F=2.5$, $D=0.5$, (b) $F=1.6$, $D=0.5$, (c) $F=0.8$, $D=0.5$. } 
	\label{FDTD}
\end{figure}

The harmonics with propagative nature appearing in \Fig{Fig4coeficientes} now scatters in different directions. The diffraction angles of each propagating harmonic have been calculated using \eqref{angle}. They have been compared with the angles obtained by FDTD in TABLE \ref{table1}. As observed, there is a good agreement between both analytical (Floquet circuit) and numerical results. Naturally, one point to note is the difference in simulation times for each solution. The analytical Floquet solution reduces notably the computational complexity compared to the FDTD.  Concretely, the circuit model requires a simulation time of the order of seconds, while the FDTD takes minutes to simulate the scenario. This becomes more evident as the macroperiod of the time-modulated metamaterial is larger.

Subsequently, \Fig{FDTD} illustrates the electric field distribution in the transmission region ($z>0$) for the cases reported in \Fig{Fig4coeficientes}. \Fig{FDTD}(a) considers $F = 2.5$ and $D = 0.5$. The time screen is located at $z/\lambda_0 = 0$.  As $F$ substantially decreases, it can be noticed that the diffraction angle of higher-order harmonics separate from that of the fundamental harmonic ($\theta_{0}=30^\mathrm{o}$), approaching the normal direction ($\theta_{n} \approx 0^\mathrm{o}$). This is well predicted by \eqref{angle}.  It is worth remarking that the harmonic orders appearing in TABLE \ref{table1} are those whose amplitude contribution is significant in the field representation of \Fig{FDTD}. The rest of harmonics taking place in the whole field expansion have not been included for some reasons: their amplitude is not significant and are not appreciated in the FDTD; they have an evanescent nature; they propagate backwards  ($\beta^{(2)}_{n}<0$). 


\begin{figure}[!t]
	\centering
        \includegraphics[width=0.5\textwidth]{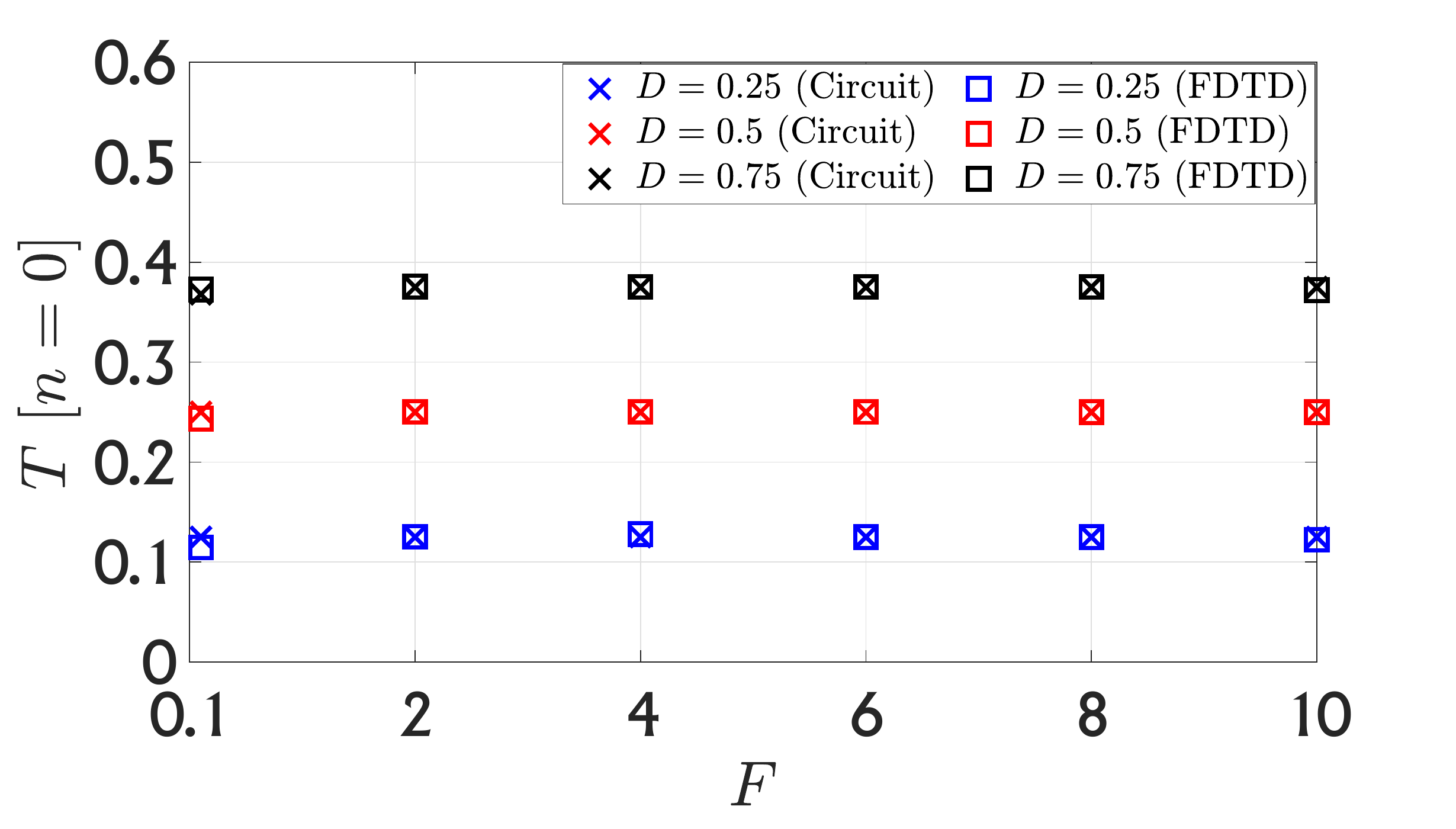}
	\caption{\small Transmission coefficient $T$ as a function of the modulation ratio $F$ for different duty cycles $D$.} 
	\label{T_F}
\end{figure}

Finally, \Fig{T_F} shows the transmission coefficient $T$, related to the fundamental harmonic ($n=0$), for several values of the modulation ratio $F$ and duty cycle $D$. Normal incidence is now considered, though oblique incidence can straightforwardly be computed A comparison is illustrated between the results extracted from the Floquet circuit and the FDTD method, showing an good agreement. It can be appreciated that, for a fixed duty cycle, the transmission coefficient remains constant regardless of the value of the modulation ratio. Conversely, the transmission coefficient increases as the duty cycle does. This is due to the fact that the time-periodic screen remains a greater amount of time in the "air" state than in the "metal" state, allowing the incident waves to pass through it more easily in average.

To conclude, in this Letter, we have studied the diffraction of electromagnetic fields produced by an incident plane wave with TE/TM polarization impinging on a time-periodic metallic screen. The proposed time-modulated metamaterial periodically alternates between "air" and "metal" states, leading to the excitation of diffraction orders that can be exploited to manipulate the propagation of electromagnetic waves. We have carried out the analysis by means of two tools: an analytical Floquet circuit and a numerical FDTD method.  By introducing the concepts of ``macroperiod" ($T_\mathrm{m}$) and "duty cycle" ($D$) to the time modulation, we have extended the beamforming capabilities of the temporal structure shown in our previous works. The reconfigurability of higher-order modes has been discussed as a function of changes in the modulation ratio $F$  and the duty cycle $D$. These results open up the possibility to simulate time-varying structures in a much more faster and efficient way than other full-wave electromagnetic tools, with the aim of designing novel time-based microwave and photonic devices.

\begin{acknowledgments}
This work was supported in part by the Spanish Government under Projects PID2020-112545RB-C54, TED2021-129938B-I00 and TED2021-131699B-I00; in part by "Junta de Andalucía” under Project A-TIC-608-UGR20, Project P18.RT.4830, and Project PYC20-RE-012-UGR; in part by a Leonardo Grant of the BBVA foundation. The authors acknowledge the support of the BBVA foundation for the funds associated to a project belonging to the program
Leonardo Grants 2021 for researchers and cultural creators
from the BBVA foundation. The BBVA Foundation accepts no responsibility for the opinions, statements and contents included in the project and/or the results thereof, which are entirely the responsibility of the authors.
\end{acknowledgments}

\section*{Data Availability}

The data that support the findings of this study (analytical Floquet circuit and numerical FDTD codes) are available from the corresponding author upon reasonable request.

\section*{References}
\nocite{*}
\bibliography{aipsamp}

\end{document}